\RequirePackage{fix-cm}
\documentclass[twocolumn,epjc3]{svjour3}  
\smartqed  
\RequirePackage{graphicx}

\usepackage{amsfonts}
\usepackage{amsmath}
\usepackage{amssymb}
\usepackage{bigints}
\usepackage{booktabs} 
\usepackage{multirow}
\usepackage{subfig}
\usepackage{lineno}
\usepackage{color}
\usepackage{comment}
\usepackage{bm}
\RequirePackage[colorlinks,citecolor=blue,urlcolor=blue,linkcolor=blue]{hyperref}
\journalname{Eur. Phys. J. C}
\newcommand{\onbb}{$0\nu\beta\beta$}

\newcommand{\Tonbb}{$T_{1/2}^{\,0\nu}$}

\newcommand{\Qbb}{$Q_{\beta\beta}$}

\newcommand{\Te}{$^{130}$Te}

\newcommand{\Co}{$^{60}$Co}
\newcommand{\ctsper}{cts$/($keV$\cdot$kg$\cdot$yr$)$}
\newcommand{\cuore}{CUORE}
\newcommand{\cuoreo}{CUORE-0}
\newcommand{\cuoricino}{Cuoricino}
\newcommand{\bi}{$BI$}

\def\fauxschelper#1 #2\relax{%
  \fauxschelphelp#1\relax\relax%
  \if\relax#2\relax\else\ \fauxschelper#2\relax\fi%
}
\def\Hscale{.85}\def\Vscale{.74}\def\Cscale{1.12}
\def\fauxschelphelp#1#2\relax{%
  \ifnum`#1>``\ifnum`#1<`\{\scalebox{\Hscale}[\Vscale]{\uppercase{#1}}\else%
    \scalebox{\Cscale}[1]{#1}\fi\else\scalebox{\Cscale}[1]{#1}\fi%
  \ifx\relax#2\relax\else\fauxschelphelp#2\relax\fi}

\definecolor{darkgreen}{rgb}{0.,0.8,0.}

\begin{document}


\title{CUORE Sensitivity to $\bm{0\nu\beta\beta}$ Decay
}

\author{
C.~Alduino\thanksref{USC} 
\and
K.~Alfonso\thanksref{UCLA} 
\and
D.~R.~Artusa\thanksref{USC,LNGS} 
\and
F.~T.~Avignone~III\thanksref{USC} 
\and
O.~Azzolini\thanksref{INFNLegnaro} 
\and
T.~I.~Banks\thanksref{BerkeleyPhys,LBNLNucSci} 
\and
G.~Bari\thanksref{INFNBologna} 
\and
J.W.~Beeman\thanksref{LBNLMatSci} 
\and
F.~Bellini\thanksref{Roma,INFNRoma} 
\and
G.~Benato\thanksref{BerkeleyPhys}
% ...
\\
}

\institute{Department of Physics and Astronomy, University of South Carolina, Columbia, SC 29208, USA\label{USC} 
\and
Department of Physics and Astronomy, University of California, Los Angeles, CA 90095, USA\label{UCLA} 
\and
INFN -- Laboratori Nazionali del Gran Sasso, Assergi (L'Aquila) I-67010, Italy\label{LNGS} 
\and
INFN -- Laboratori Nazionali di Legnaro, Legnaro (Padova) I-35020, Italy\label{INFNLegnaro} 
\and
Department of Physics, University of California, Berkeley, CA 94720, USA\label{BerkeleyPhys} 
\and
Nuclear Science Division, Lawrence Berkeley National Laboratory, Berkeley, CA 94720, USA\label{LBNLNucSci} 
\and
INFN -- Sezione di Bologna, Bologna I-40127, Italy\label{INFNBologna} 
\and
Materials Science Division, Lawrence Berkeley National Laboratory, Berkeley, CA 94720, USA\label{LBNLMatSci} 
\and
Dipartimento di Fisica, Sapienza Universit\`{a} di Roma, Roma I-00185, Italy\label{Roma} 
\and
INFN -- Sezione di Roma, Roma I-00185, Italy\label{INFNRoma} 
% ...
} 

\thankstext{fn2}{Deceased}
\thankstext{fn3}{Presently at: INFN -- Laboratori Nazionali di Frascati, Frascati (Roma) I-00044, Italy}





\date{Received: date / Accepted: date}

\maketitle

\begin{abstract}
  We report a study of the CUORE sensitivity to neutrinoless double beta (\onbb) decay.
  We used a Bayesian analysis based on a toy Monte Carlo (MC) approach
  to extract the exclusion sensitivity to the  \onbb\ decay half-life (\Tonbb)
  at $90\%$~credibility interval (CI) --
  i.e. the interval containing the true value of \Tonbb\ with $90\%$ probability --
  and the $3~\sigma$ discovery sensitivity.
  We consider various background levels and energy resolutions,
  and describe the influence of the data division in subsets with different background levels.

  If the background level and the energy resolution meet the expectation,
  CUORE will reach a $90\%$~CI exclusion sensitivity of $2\cdot10^{25}$~yr
  with $3$ months, and $9\cdot10^{25}$~yr with $5$~years of live time.
  Under the same conditions, the discovery sensitivity after $3$~months and $5$~years
  will be $7\cdot10^{24}$~yr and $4\cdot10^{25}$~yr, respectively.
\end{abstract}

\section{Introduction}\label{sec:Introduction}

Neutrinoless double beta decay is a non Standard Model process
that violates the total lepton number conservation
and implies a Majorana neutrino mass component~\cite{Schechter:1981bd,Duerr:2011zd}.
This decay is currently being investigated with a variety of double beta decaying isotopes.
A recent review can be found in Ref.~\cite{Dell'Oro:2016dbc}.
The Cryogenic Underground Observatory for Rare Events
(CUORE)~\cite{Artusa:2014lgv,CUORE-NIM2004,Arnaboldi:2003tu}
is an experiment searching for \onbb\ decay in \Te.
It is located at the Laboratori Nazionali del Gran Sasso of INFN, Italy.
In \cuore, $988$ TeO$_2$ crystals with natural \Te\ isotopic abundance and a $750$~g average mass
are operated simultaneously as source and bolometric detector for the decay.
In this way, the \onbb\ decay signature is a peak at the $Q$-value of the reaction
(\Qbb, $2527.518$~keV for \Te~\cite{Redshaw:2009cf,Scielzo:2009nh,Rahaman:2011zz}).
Bolometric crystals are characterized by an excellent energy resolution ($\sim0.2\%$ Full Width at Half Maximum, FWHM)
and a very low background  at \Qbb, which is expected to be at the $10^{\mbox{-}2}$~\ctsper\ level
in \cuore~\cite{Alduino:2017qet}.

The current best limit on \onbb\ decay in \Te\ comes from a combined analysis
of the \cuoreo~\cite{Alduino:2016vjd,Aguirre:2014lua}
and \cuoricino\ data~\cite{Arnaboldi:2008ds,Andreotti:2010vj}.
With a total exposure of $29.6$ kg$\cdot$yr, a limit of $T_{1/2}^{0\nu}>4.0\cdot10^{24}$~yr~($90\%$~CI)
is obtained~\cite{Alfonso:2015wkao} for the \onbb\ decay half life, \Tonbb.

After the installation of the detector, successfully completed in the summer 2016,
\cuore\ started the commissioning phase at the beginning of 2017.
The knowledge of the discovery and exclusion sensitivity to \onbb\ decay
as a function of the measurement live time can be exploited
to set the criteria for the unblinding of the data
and the release of the \onbb\ decay analysis results.

In this work, we dedicate our attention to those factors
which could strongly affect the sensitivity, such as the Background Index (\bi)
and the energy resolution at \Qbb.
In \cuore, the crystals in the outer part of the array are expected to show
a higher \bi\ than those in the middle~\cite{Alduino:2017qet}.
Considering this and following the strategy already implemented
by the \textsc{Gerda} Collaboration~\cite{Agostini:2013mzu,Agostini:2017iyd},
we show how the division of the data into subsets
with different \bi\ could improve the sensitivity.

The reported results are obtained by means of a Bayesian analysis
performed with the Bayesian Analysis Toolkit (BAT)~\cite{Caldwell:2008fw}.
The analysis is based on a toy-MC approach.
At a cost of a much longer computation time with respect to the use
of the median sensitivity formula~\cite{Alessandria:2011rc},
this provides the full sensitivity probability distribution and not only its median value.

In Section~\ref{sec:StatisticalMethod}, we review the statistical methods for the parameter estimation,
as well as for the extraction of the exclusion and discovery sensitivity.
Section~\ref{sec:ExperimentalParameters} describes the experimental parameters
used for the analysis while its technical implementation is summarized in Section~\ref{sec:FitProcedure}.
Finally, we present the results in Section~\ref{sec:FitResults}.

\section{Statistical Method}\label{sec:StatisticalMethod}

The computation of exclusion and discovery sensitivities presented here
follows a Bayesian approach: we exploit the Bayes theorem both for parameter estimation
and model comparison.
In this work, we use the following notation:
\begin{itemize}
\item $H$ indicates both a hypothesis and the corresponding model;
\item $H_0$ is the background-only hypothesis, according to which the known physics processes
  are enough to explain the experimental data.
  In the present case, we expect the \cuore\ background to be flat in a $100$~keV region around \Qbb,
  except for the presence of a \Co\ summation peak at $2505.7$~keV.
  Therefore, $H_0$ is implemented as a flat background distribution
  plus a Gaussian describing the \Co\ peak. In \cuoreo, this peak was found to be centered
  at an energy $1.9\pm0.7$~keV higher than that tabulated in literature~\cite{Alfonso:2015wkao}.
  This effect, present also in \cuoricino~\cite{Andreotti:2010vj}, is a feature of all gamma summation peaks.
  Hence, we will consider the \Co\ peak to be at $2507.6$~keV.
\item $H_1$ is the background-plus-signal hypothesis,
  for which some new physics is required to explain the data.
  In our case, the physics involved in $H_1$ contains the background processes as well as \onbb\ decay.
  The latter is modeled as a Gaussian peak at \Qbb.
\item $\vec{E}$ represents the data. It is a list of $N$ energy bins
  centered at the energy $E_i$ and containing $n_i$ event counts.
  The energy range is $[2470;2570]$~keV.
  This is the same range used for the \cuoreo\ \onbb\ decay analysis~\cite{Alfonso:2015wkao},
  and is bounded by the possible presence of peaks from $^{214}$Bi at $2447.7$~keV
  and $^{208}$Tl X-ray escape at $\sim 2585$~keV~\cite{Alfonso:2015wkao}.
  While an unbinned fit allows to fully exploit the information contained in the data,
  it can result in a long computation time for large data samples.
  Given an energy resolution of $\sim5$~keV FWHM and using a $1$~keV bin width,
  the $\pm3$~sigma range of a Gaussian peak is contained in $12.7$~bins.
  With the $1$~keV binning choice, the loss of information with respect to the unbinned fit is negligible.
\item $\Gamma^{0\nu}$ is the parameter describing the \onbb\ decay rate for $H_1$:
  \begin{equation}
    \Gamma^{0\nu} = \frac{\ln{2}}{T_{1/2}^{0\nu}}\ .
  \end{equation}
\item $\vec{\theta}$ is the list of nuisance parameters describing the background processes
  in both $H_0$ and $H_1$;
\item $\Omega$ is the parameter space for the parameters $\vec{\theta}$.
\end{itemize}

\subsection{Parameter Estimation}\label{sec:ParameterEstimation}

We perform the parameter estimation for a model $H$ through the Bayes theorem,
which yields the probability distribution for the parameters based on the measured data,
under the assumption that the model $H$ is correct.
In the \onbb\ decay analysis, we are interested in the measurement of $\Gamma^{0\nu}$ for the hypothesis $H_1$.
The probability distribution for the parameter set $(\Gamma^{0\nu},\vec{\theta})$ is:
\begin{multline}\label{eq:BayesParameter}
  P\left(\Gamma^{0\nu},\vec{\theta} \big| \vec{E},H_1\right) =\\
  \frac{ P\left( \vec{E} \big| \Gamma^{0\nu}, \vec{\theta}, H_1 \right)
    \pi(\Gamma^{0\nu})  \pi(\vec{\theta}) }
       { \bigintsss_{\Omega}\bigintsss_0^{\infty}
         P\left( \vec{E} \big| \Gamma^{0\nu}, \vec{\theta}, H_1 \right)
         \pi(\Gamma^{0\nu})  \pi(\vec{\theta})\, d\vec{\theta}\, d\Gamma^{0\nu}
       }\ .
\end{multline}
The numerator contains the conditional probability\\ $P\left( \vec{E} \big| \Gamma^{0\nu}, \vec{\theta}, H_1 \right)$
of finding the measured data $\vec{E}$ given the model $H_1$ for a set of parameters $(\Gamma^{0\nu},\vec{\theta})$,
times the prior probability $\pi$ for each of the considered parameters.
The prior probability has to be chosen according to the knowledge
available before the analysis of the current data.
For instance, the prior for the number of signal counts $\Gamma^{0\nu}$ might be based
on the half-life limits reported by previous experiments
while the prior for the background level in the region of interest (ROI)
could be set based on the extrapolation of the background measured outside the ROI.
The denominator represents the overall probability to obtain the data $\vec{E}$
given the hypothesis $H_1$ and all possible parameter combinations, $P(\vec{E}|H_1)$.

The posterior probability distribution for $\Gamma^{0\nu}$ is obtained via marginalization,
i.e. integrating\\ $P\left(\Gamma^{0\nu},\vec{\theta} \big| \vec{E},H_1\right)$
over all nuisance parameters $\vec{\theta}$:
\begin{equation}\label{eq:posterior}
  P\left( \Gamma^{0\nu} \big| H_1, \vec{E} \right) =
  \bigintsss_\Omega P\left(\Gamma^{0\nu},\vec{\theta} \big| \vec{E},H_1\right) d\vec{\theta}\ .
\end{equation}

For each model $H$, the probability of the data given the model and the parameters has to be defined.
For a fixed set of experimental data, this corresponds to the likelihood function~\cite{james}.
Dividing the data into $N_d$ subsets with index $d$ characterized by different background levels,
and considering a binned energy spectrum with $N$ bins
and a number $n_{di}$ of events in the bin $i$ of the $d$ subset spectrum,
the likelihood function is expressed by the product of a Poisson term for each bin $di$:
\begin{multline}\label{eq:binnedL}
  P\left( \vec{E} \big| \Gamma^{0\nu},\vec{\theta}, H \right)
   = \mathcal{L}\left( \vec{E} \big| \Gamma^{0\nu},\vec{\theta}, H \right)\\
   = \prod_{d=1}^{N_d}\  \prod_{i=1}^{N}
  \frac{ e^{-\lambda_{di}} \cdot \lambda_{di}^{n_{di}} }{ n_{di}! } \ ,
\end{multline}
where $\lambda_{di}$ is the expectation value for the bin $di$.
The best-fit is defined as the set of parameter values $(\Gamma^{0\nu},\vec{\theta})$
for which the likelihood is at its global maximum.
In the practical case, we perform the maximization on the log-likelihood
\begin{equation}\label{eq:binnedLogL}
  \ln{ \mathcal{L}\left( \vec{E} \big| \Gamma^{0\nu}, \vec{\theta}, H \right) }
  = \sum_{d=1}^{N_d} \ \sum_{i=1}^{N} \left( -\lambda_{di} + \ln{\lambda_{di}^{n_{di}}} \right)\ ,
\end{equation}
where the additive terms $-\ln{(n_{di}!)}$ are dropped from the calculation.

The difference between $H_0$ and $H_1$ is manifested in the formulation of $\lambda_{di}$.
As mentioned above, we parametrize $H_0$ with a flat distribution over
the considered energy range, i.e. $[2470;2570]$~keV:
\begin{equation}\label{eq:bkgpdf}
  f_{bkg}(E) = \frac{1}{E_{max} - E_{min}}
\end{equation}
plus a Gaussian distribution for the \Co\ peak:
\begin{equation}\label{eq:60Copdf}
  f_{Co}(E) =   \frac{1}{\sqrt{2\pi}\ \sigma}
  \exp{\left( - \frac{\left(E-\mu_{Co}\right)^2}{2\sigma^2} \right)}\ .
\end{equation}
The expected background counts in the bin $di$
corresponds to the integral of $f_{bkg}(E)$ in the bin $di$ times the total number
of background counts $M^{bkg}_d$ for the subset $d$:
\begin{equation}\label{eq:bkgLambda}
  \lambda^{bkg}_{di} 
    = \bigintsss_{E_{di}^{min}}^{E_{di}^{max}} M^{bkg}_d f_{bkg}(E) dE
\end{equation}
where $E^{min}_{di}$ and $E^{max}_{di}$ are the left and right margins of the energy bin $di$, respectively.
Considering bins of size $\delta E_{di}$ and expressing $M^{bkg}_{di}$
as function of the background index $BI_d$, of the total mass $m_d$
and of the measurement live time $t_d$, we obtain:
\begin{equation}\label{eq:bkgLambda2}
  \lambda^{bkg}_{di} = BI_d \cdot m_d \cdot t_d \cdot \delta E_{di}\ .
\end{equation}

Similarly, the expectation value for the \Co\ distribution on the bin $di$ is:
\begin{equation}\label{eq:60CoLambda}
  \lambda_{di}^{Co} = \bigintss_{E^{min}_{di}}^{E^{max}_{di}}
  \frac{M^{Co}_d}{\sqrt{2\pi}\ \sigma}
  \exp{\left( - \frac{\left(E-\mu_{Co}\right)^2}{2\sigma^2} \right)}dE\ ,
\end{equation}
where $M^{Co}_d$ is the total number of \Co\ events for the subset $d$
and can be redefined as function of the \Co\ event rate, $R^{Co}_d$:
\begin{equation}
  M^{Co}_d = R^{Co}_d \cdot m_d \cdot t_d\ .
\end{equation}
The total expectation value $\lambda_{di}$ for $H_0$ is then:
\begin{equation}
  \lambda_{di} = \lambda^{bkg}_{di} + \lambda^{Co}_{di}\ .
\end{equation}

In the case of $H_1$ an additional expectation value for \onbb\ decay is required:
\begin{equation}\label{eq:onbbLambda}
  \lambda_{di}^{0\nu} = \bigintss_{E^{min}_{di}}^{E^{max}_{di}}
  \frac{M^{0\nu}_d}{\sqrt{2\pi}\ \sigma}
  \exp{\left( - \frac{\left(E-Q_{\beta\beta}\right)^2}{2\sigma^2} \right)} dE\ .
\end{equation}
The number of \onbb\ decay events in the subset $d$ is:
\begin{equation}\label{eq:SgnCountsToHalflife}
  M^{0\nu}_d = \Gamma^{0\nu} \frac{ N_A }{ m_A } \cdot f_{130} \cdot \varepsilon_{tot} \cdot m_d \cdot t_d\ ,
\end{equation}
where $N_A$ is the Avogadro number, $m_a$ and $f_{130}$ are the molar mass and the isotopic abundance of \Te\ and
$\varepsilon_{tot}$ is the total efficiency, i.e. the product of the containment efficiency $\varepsilon_{MC}$
(obtained with MC simulations) and the instrumental efficiency $\varepsilon_{instr}$.

\subsection{Exclusion Sensitivity}\label{sec:ExclusionSensitivity}

\begin{figure}[b]
  \def\svgwidth{\columnwidth}
  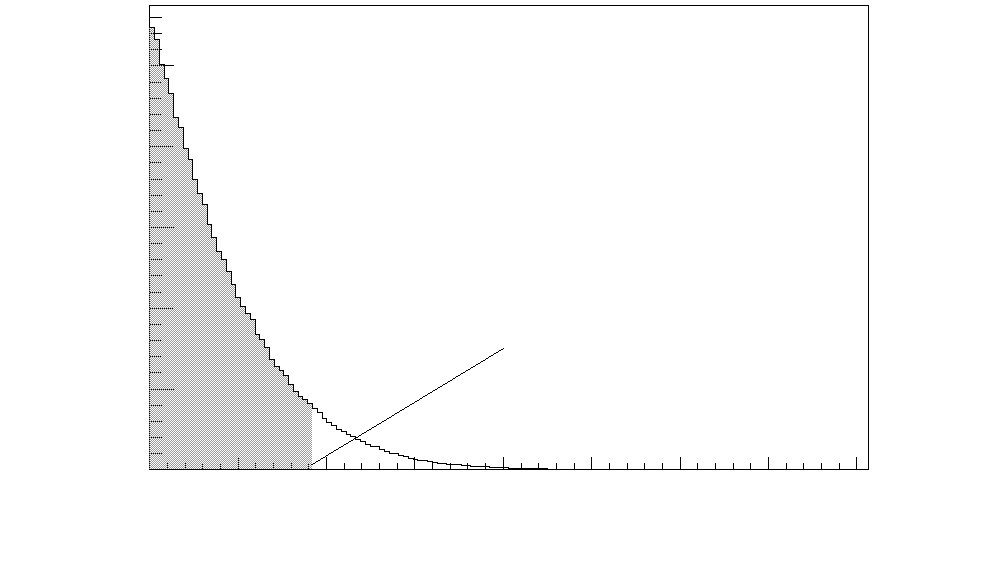\\
  \ \\
  \def\svgwidth{\columnwidth}
  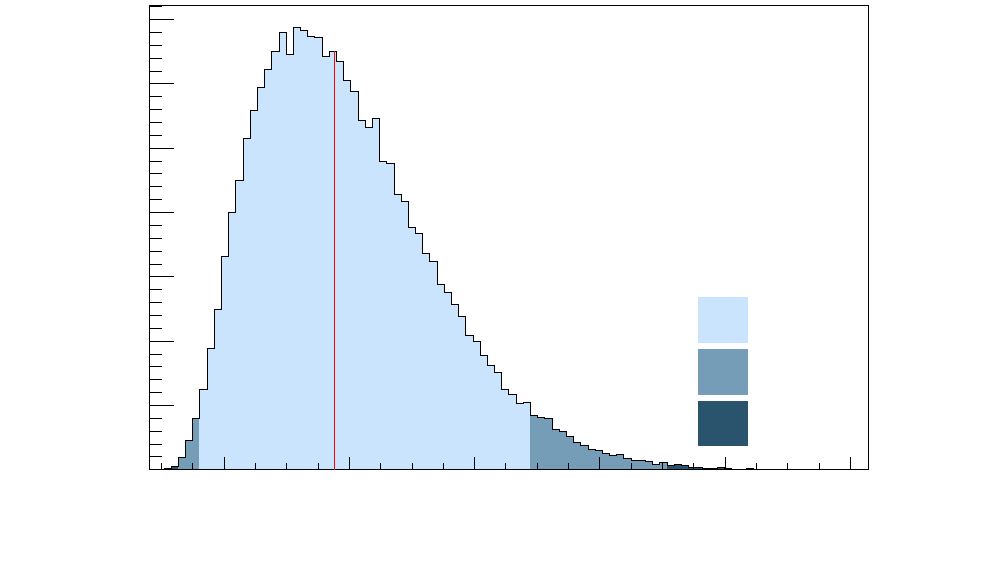
  \caption{\small{Top: marginalized probability distribution for $\Gamma^{0\nu}$ relative to one toy-MC experiment.
    The $90\%$~CI limit on \Tonbb\ is indicated.
    Bottom: distribution of the $90\%$~CI limits on $T^{0\nu}_{1/2}$ for $10^5$ toy-MC experiments.
    The red vertical line corresponds to the median sensitivity, while the three colors depict
    the $1$, $2$ and $3~\sigma$ quantiles of the distribution.
    Both plots are obtained with $1$~yr live time, a $10^{\mbox{-}2}$~\ctsper\ $BI$ and $5$~keV energy resolution.
    }}
  \label{fig:marginalized}
\end{figure}

We compute the exclusion sensitivity by means of the $90\%$~CI limit.
This is defined as the value of \Tonbb\ corresponding to the $90\%$ quantile
of the posterior $\Gamma^{0\nu}$ distribution:
\begin{multline}\label{eq:limit}
  T_{1/2}^{\,0\nu}\left( 90\%\ CI \right) = T_{1/2}^{\,0\nu} : \\
  \bigintsss_{0}^{\ln{2}/T_{1/2}^{\,0\nu}}
  P\left( \Gamma^{0\nu} \big| H_1, \vec{E} \right) d\Gamma^{0\nu} = 0.9\ .
\end{multline}
An example of posterior probability for $\Gamma^{0\nu}$ and the relative $90\%$~CI limit
is shown in Fig.~\ref{fig:marginalized}, top. Flat prior distributions are used for all parameters,
as described in Sec.~\ref{sec:ExperimentalParameters}.

In the Bayesian approach, the limit is a statement regarding the true value of the considered physical quantity.
In our case, a $90\%$~CI limit on \Tonbb\ is to be interpreted as the value above which,
given the current knowledge, the true value of \Tonbb\ lies with $90\%$ probability.
This differs from a frequentist $90\%$~C.L. limit, which is a statement regarding the possible results
of the repetition of identical measurements and should be interpreted as the value
above which the best-fit value of \Tonbb\ would lie in the $90\%$ of the imaginary identical experiments.

In order to extract the exclusion sensitivity, we generate a set of $N$ toy-MC spectra
according to the back-ground-only model, $H_0$.
We then fit spectra with the background-plus-signal model, $H_1$,
and obtain the\\ $T_{1/2}^{\,0\nu}\left( 90\%\ CI \right)$ distribution (Fig.~\ref{fig:marginalized}, bottom).
Its median $\hat{T}_{1/2}^{\,0\nu}\left( 90\%\ CI \right)$ is referred as the median sensitivity.
For a real experiment, the experimental \Tonbb\ limit is expected to be above/below
$\hat{T}_{1/2}^{\,0\nu}\left( 90\%\ CI \right)$ with $50\%$ probability.
Alternatively, one can consider the mode of the distribution,
which corresponds to the most probable \Tonbb\ limit.

The exact procedure for the computation of the exclusion sensitivity is the following:
\begin{itemize}
\item for each subset, we generate a random number of background events $N_d^{bkg}$
  according to a Poisson distribution with mean $\lambda^{bkg}_d$;
\item for each subset, we generate $N_d^{bkg}$ events with an energy randomly distributed
  according to $f_{bkg}(E)$;
\item we repeate the procedure for the \Co\ contribution;
\item we fit the toy-MC spectrum with the $H_1$ model (Eq.~\ref{eq:BayesParameter}),
  and marginalize the likelihood with respect to the parameters $BI_d$ and $R_d^{Co}$ (Eq.~\ref{eq:posterior});
\item we extract the $90\%$~CI limit on \Tonbb;
\item we repeat the algorithm for $N$ toy-MC experiments, and build the distribution of
  $T_{1/2}^{\,0\nu}\left( 90\%\ CI \right)$.
\end{itemize}

\subsection{Discovery Sensitivity}\label{sec:DiscoverySensitivity}

The discovery sensitivity provides information on the required strength of the signal amplitude
for claiming that the known processes alone are not sufficient to properly
describe the experimental data.
It is computed on the basis of the comparison between the background-only
and the background-plus-signal models. A method for the calculation of the Bayesian
discovery sensitivity was introduced in Ref.~\cite{Caldwell:2006yj}.
We report it here for completeness.

In our case, we assume that $H_0$ and $H_1$ are a complete set of models, for which:
\begin{equation}
  P\bigl(H_0\big|\vec{E}\bigr) + P\bigl(H_1\big|\vec{E}\bigr) = 1\ .
\end{equation}
The application of the Bayes theorem to the models $H_0$ and $H_1$ yields:
\begin{align}\label{eq:BayesModels}
  \begin{split}
  P\bigl(H_0\big|\vec{E}\bigr) & = \frac{ P\bigl( \vec{E}\big|H_0 \bigr) \pi(H_0) }{ P\bigl(\vec{E}\bigr) } \\
  P\bigl(H_1\big|\vec{E}\bigr) & = \frac{ P\bigl( \vec{E}\big|H_1 \bigr) \pi(H_1) }{ P\bigl(\vec{E}\bigr) }\ .
  \end{split}
\end{align}
In this case, the numerator contains the probability of measuring the data $\vec{E}$ given the model $H$:
\begin{multline}
  \begin{split}
  P\bigl( \vec{E}\big|H_0 \bigr)
  = \bigintsss_{\Omega}
  P\left( \vec{E} \big| \vec{\theta}, H_0 \right)
  \pi(\vec{\theta})\, d\vec{\theta}\\
  P\bigl( \vec{E}\big|H_1 \bigr) 
  = \bigintsss_{\Omega}\bigintsss_0^{\infty}
  P\left( \vec{E} \big| \Gamma^{0\nu}, \vec{\theta}, H_1 \right)\times\quad\quad\\
  \times\pi(\Gamma^{0\nu})  \pi(\vec{\theta})\, d\vec{\theta}\, d\Gamma^{0\nu}\ ,
  \end{split}
\end{multline}
while the prior probabilities for the models $H_0$ and $H_1$ can be chosen
as $0.5$ so that neither model is favored.

The denominator of Eq.~\ref{eq:BayesModels} is the sum probability
of obtaining the data $\vec{E}$ given either the model $H_0$ or $H_1$:
\begin{equation}\label{eq:ProbData}
  P\bigl(\vec{E}\bigr)
  =  P\bigl( \vec{E}\big|H_0 \bigr) \pi(H_0) + P\bigl( \vec{E}\big|H_1 \bigr) \pi(H_1)\ .
\end{equation}

At this point we need to define a criterion for claiming the discovery of new physics.
Our choice is to quote the $3~\sigma$ (median) discovery sensitivity,
i.e. the value of \Tonbb\ for which the posterior probability
of the back-ground-only model $H_0$ given the data is smaller than $0.0027$ in $50\%$
of the possible experiments. In other words:
\begin{multline}\label{eq:discoverySensitivity}
  \hat{T}_{1/2}^{0\nu}(3\sigma) =
  T_{1/2}^{0\nu} : P\bigl( H_0 \big| \vec{E} \bigr) < 0.0027\\ \text{~for~}N/2~\text{experiments.}
\end{multline}

The detailed procedure for the determination of the discovery sensitivity is:
\begin{itemize}
\item we produce a toy-MC spectrum according to the $H_1$ model
  with an arbitrary value of \Tonbb;
\item we fit the spectrum with both $H_0$ and $H_1$;
\item we compute $P(H_0 | \vec{E})$;
\item we repeat the procedure for $N$ toy-MC spectra using the same \Tonbb;
\item we repeat the routine with different values of \Tonbb\
  until the condition of Eq.~\ref{eq:discoverySensitivity} is satisfied.
  The iteration is implemented using the bisection method
  until a $5\cdot10^{\mbox{-}5}$ precision is obtained on the median $P(H_0 | \vec{E})$.
\end{itemize}

\section{Experimental Parameters}\label{sec:ExperimentalParameters}

The fit parameters of the $H_1$ model are \bi, $R^{Co}$ and $\Gamma^{0\nu}$,
while only the first two are present for $H_0$.
If the data are divided in subsets,
different \bi\ and $R^{Co}$ fit parameter are considered for each subset.
On the contrary, the inverse \onbb\ half-life is common to all subsets.

Prior to the assembly of the \cuore\ crystal towers, we performed a screening survey of the employed
materials~\cite{Alessandria:2011vj,Barghouty:2010kj,Wang:2015pxa,Alessandria:2012zp,Andreotti:2009dk,Bellini:2009zw,Andreotti:2009zza,giachero}.
From these measurements, either a non-zero activity was obtained,
or a $90\%$ confidence level (C.L.) upper limit was set.
Additionally, the radioactive contamination of the crystals and holders was also obtained
from the \cuoreo\ background model~\cite{Alduino:2016vtd}.
We developed a full MC simulation of \cuore~\cite{Alduino:2017qet},
and we used the results of the screening measurements and of the \cuoreo\ background model
for the normalization of the simulated spectra.
We then computed the \bi\ at \Qbb\ using the output of the simulations.
In the present study, we consider only those background contributions
for which a non-zero activity is obtained from the available measurements.
The largest background consists of $\alpha$ particles emitted by U and Th surface contaminations
of the copper structure holding the crystals.
Additionally, we consider a \Co\ contribution normalized to the $90\%$~C.L. limit from the screening measurement.
In this sense, the effect of a \Co\ background on the \cuore\ sensitivity is to be held as an upper limit.
Given the \Co\ importance especially in case of sub-optimal energy resolution,
we preferred to maintain a conservative approach in this regard.
In the generation of the toy-MC spectra, we take into account the  \Co\ half life ($5.27$~yr),
and set the start of data taking to January $2017$.

The parameter values used for the production of the toy-MC are reported
in Tab.~\ref{tab:parameters}.
The quoted uncertainty on the $BI$ comes from the CUORE MC simulations~\cite{Alduino:2017qet}.
We produce the toy-MC spectra using the best-fit value of the $BI$.
In a second time, we repeat the analysis after increasing and decreasing the $BI$
by an amount equivalent to its statistical and systematic uncertainties combined in quadrature.

\begin{table}[t]
  \caption{\small{Input parameters for the production of toy-MC spectra.}}
  \label{tab:parameters}
  \begin{tabular}{cc}
    \toprule
    \bi\ [cts$/($keV$\cdot$kg$\cdot$yr$)$] & $R^{Co}$ [cts$/($kg$\cdot$yr$)$] \\
    \midrule
    $\left(1.02\pm0.03(\text{stat})^{+0.23}_{-0.10}(\text{syst})\right)\cdot10^{\mbox{-}2}$ & $0.428$ \\
    \bottomrule
  \end{tabular}
\end{table}

After running the fit on the entire crystal array as if it were a unique detector,
we considered the possibility of dividing the data grouping the crystals with a similar \bi.
Namely, being the background at \Qbb\ dominated by surface $\alpha$ contamination of the copper structure,
the crystals facing a larger copper surface are expected to have a larger \bi.
This effect was already observed in \cuoreo, where the crystals in the uppermost and lowermost levels,
which had $3$ sides facing the copper shield,
were characterized by a larger background than those in all other levels,
which were exposed to coppper only on $2$ sides.
Considering the \cuore\ geometry, the crystals can be divided in $4$ subsets
with different numbers of exposed faces. Correspondingly, they are
characterized by different \bi, as reported in Tab.~\ref{tab:subsets}.

A major ingredient of a Bayesian analysis is the choice of the priors.
In the present case, we use a flat prior for all parameters.
In particular, the prior distribution for $\Gamma^{0\nu}$ is flat between zero
and a value large enough to contain $>99.999\%$ of its posterior distribution.
This corresponds to the most conservative choice.
Any other reasonable prior, e.g. a scale invariant prior on $\Gamma^{0\nu}$, would yield a stronger limit.
A different prior choice based on the real characteristic of the experimental spectra
might be more appropriate for \bi\ and $R^{Co}$ in the analysis of the \cuore\ data.
For the time being the lack of data prevents the use of informative priors.
As a cross-check, we performed the analysis using the \bi\ and \Co~rate uncertainties
obtained by the background budget as the $\sigma$ of a Gaussian prior.
No significant difference was found on the sensitivity band
because the Poisson fluctuations of the generated number of background and \Co\ events
are dominant for the extraction of the $\Gamma^{0\nu}$ posterior probability distribution.

Tab.~\ref{tab:constants} lists the constant quantities present in the formulation
of $H_0$ and $H_1$. All of them are fixed,
with the exception of the live time $t$ and the FWHM of the \onbb\ decay and \Co\ Gaussian peaks.
We perform the analysis with a FWHM of $5$ and $10$~keV,
corresponding to a $\sigma$ of $2.12$ and $4.25$~keV, respectively.
Regarding the efficiency, while in the toy-MC production the $BI$ and $R^{Co}$
are multiplied by the instrumental efficiency\footnote{The containment efficiency is already
  encompassed in $BI$ and $R^{Co}$~\cite{Alduino:2017qet}.},
in the fit the total efficiency is used.
This is the product of the containment and instrumental efficiency.
Also in this case, we use the same value as for \cuoreo, i.e. $81.3\%$~\cite{Alfonso:2015wkao}.
We evaluate the exclusion and discovery sensitivities for different live times,
with $t$ ranging from $0.1$ to $5$~yr and using logarithmically increasing values: $t_{i} = 1.05\cdot t_{i-1}$.

\begin{table}[h]
  \caption{\small{Crystal subsets with different expected $\alpha$ background in \cuore.
    The values of $BI$ and $R^{Co}$ are taken from~\cite{Alduino:2017qet}.}}
  \label{tab:subsets}
  \begin{tabular}{lcccc}
    \toprule
    Subset & Free  & Number of &\bi~[cts$/$ & $R^{Co}$~[cts \\
    Name   & Sides & crystals & $($keV$\cdot$kg$\cdot$yr$)$] & $/($kg$\cdot$yr$)$] \\
    \midrule
    Inner    & $0$ & $528$ & $0.82(2)\cdot10^{\mbox{-}2}$ & $0.40$ \\
    Middle-1 & $1$ & $272$ & $1.17(4)\cdot10^{\mbox{-}2}$ & $0.47$ \\
    Middle-2 & $2$ & $164$ & $1.36(4)\cdot10^{\mbox{-}2}$ & $0.43$ \\
    Outer    & $3$ & $24$  & $1.78(7)\cdot10^{\mbox{-}2}$ & $0.59$ \\
    \bottomrule
  \end{tabular}
\end{table}

\begin{table}[h]
  \caption{\small{Constants used in $H_0$ and $H_1$.}}
  \label{tab:constants}
  \begin{tabular}{ccc}
    \toprule
    Constant & Symbol & Value \\
    \midrule
    Detector Mass       & $m_d$ & $741.67$~kg\\[2pt]
    Avogadro number & $N_A$ & $6.022\cdot10^{23}$~mol$^{-1}$\\[2pt]
    Molar mass & $m_A$ & $159.6$~g$/$mol\\[2pt]
    Live Time  & $t_d$ & $0.1$--$5$~yr\\[2pt]
    Efficiency & $\varepsilon_{tot}$ & $81.3\%$\\[2pt]
    \Te\ abundance & $f_{130}$ & $0.34167$\\[2pt]
    \onbb\ Q-value & \Qbb & $2527.518$~keV \\[2pt]
    \Co\ peak position & $\mu_{Co}$ & $(2505.692+1.9)$~keV\\[2pt]
   Energy resolution & FWHM & $5,10$~keV \\
    \bottomrule
  \end{tabular}
\end{table}

\section{Fit Procedure}\label{sec:FitProcedure}

We perform the analysis with the software BAT v1.1.0-DEV~\cite{Caldwell:2006yj}, which internally uses
CUBA~\cite{Hahn:2004fe} v4.2 for the integration of multi-dimensional probabilities
and the Metropolis-Hastings algorithm~\cite{Metropolis} for the fit.
The computation time depends on the number of samples drawn from the considered probability distribution.
For the exclusion sensitivity, we draw $10^5$ likelihood samples for every toy-MC experiment,
while, due to the higher computational cost, we use only $10^3$ for the discovery sensitivity.
For every combination of live time, \bi\ and energy resolution,
we run $10^5$ ($10^3$) toy-MC experiments for the exclusion (discovery) sensitivity study.
In the case of the discovery sensitivity, we chose
the number of toy-MC experiments as the minimum for which
a $2\%$ relative precision was achievable on the median sensitivity.
For the exclusion sensitivity, it was possible to increase both the number of toy-MC experiments
and iterations, with a systematic uncertainty on the median sensitivity
at the per mil level.

\section{Results and Discussion}\label{sec:FitResults}

\subsection{Exclusion Sensitivity}\label{sec:ExclusionSensitivityResults}

\begin{figure*}
  \centering
  \subfloat{
    \def\svgwidth{0.45\textwidth}
    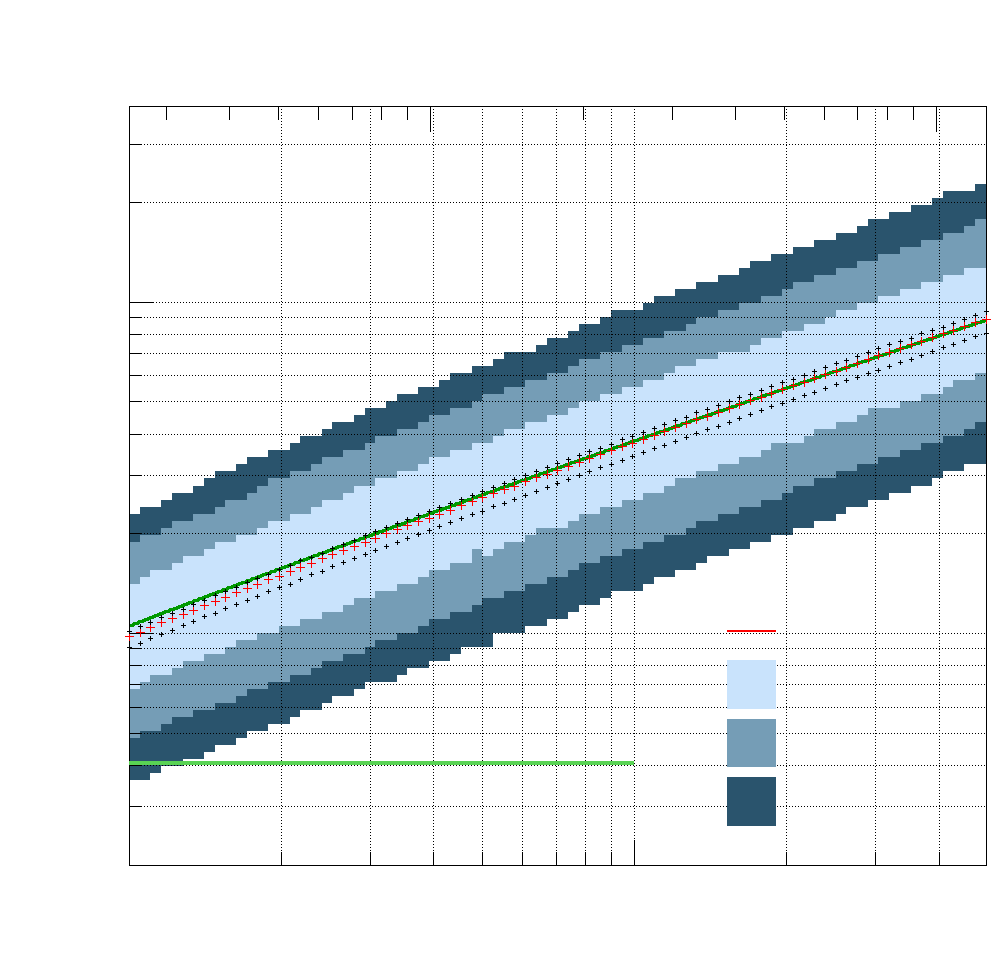} \quad
  \subfloat{
    \def\svgwidth{0.45\textwidth}
    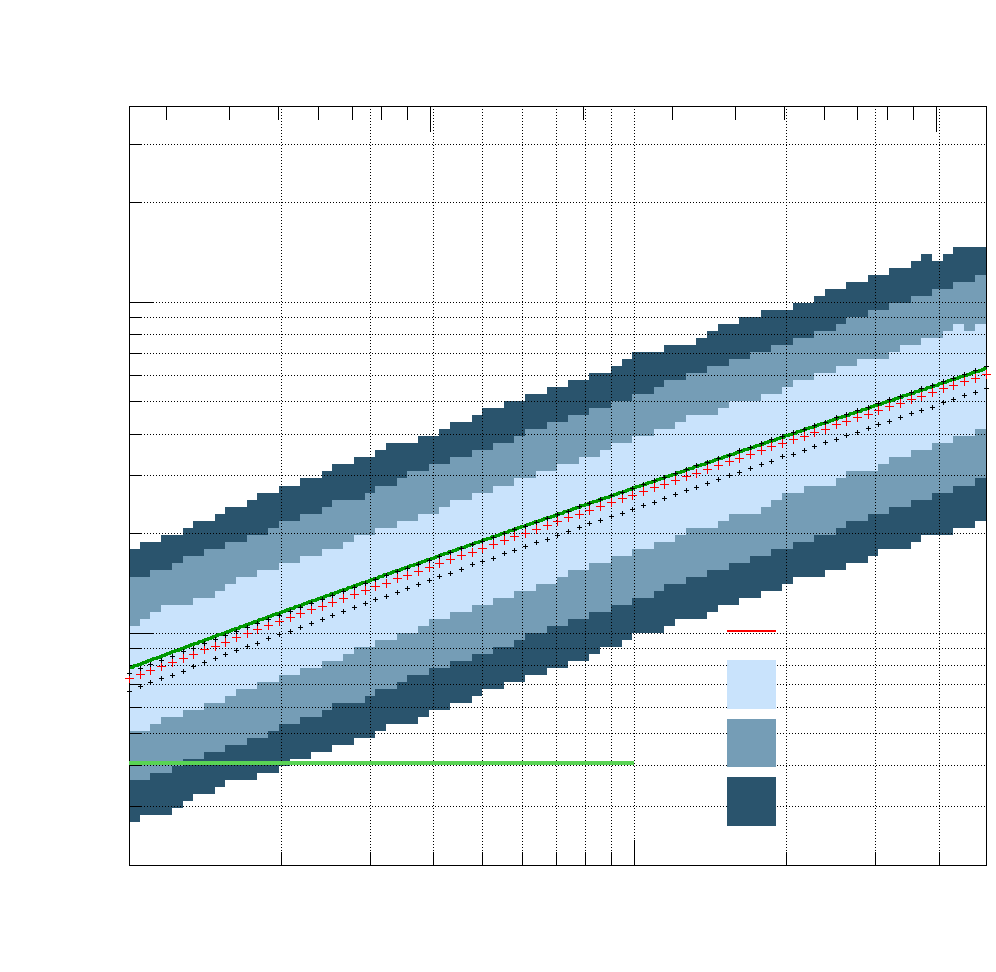} \\
  \caption{\small{$90\%$~CI exclusion sensitivity for a $5$~keV (left) and a $10$~keV FWHM (right).
    The red crosses correspond to the median sensitivity for a $BI$ of $10^{\mbox{-}2}$~\ctsper,
    while the smaller black crosses correspond to the median sensitivity obtained after shifting the $BI$
    up and down by an amount equivalent to its statistical and systematic uncertainty summed in quadrature.
    The uncertainty on $BI$ yields a $^{+4\%}_{-9\%}$ effect on the median sensitivity.
    The different colored areas depict the ranges containing the
    $68.3$, $95.5$ and $99.7\%$ of the toy-MC experiments.
    They are computed for each live time value separately as described
    in Fig.~\ref{fig:marginalized}, bottom.
    We also show the sensitivity computed as in~\cite{Alessandria:2011rc} in dark green.
    The horizontal green line at $4\cdot10^{24}$~yr corresponds to the limit obtained
    with \cuoreo\ and \cuoricino~\cite{Alfonso:2015wkao}.}}
  \label{fig:ExclusionSensitivity}
\end{figure*}

The distributions of $90\%$~CI limit as a function of live time with no data subdivision
are shown in Fig.~\ref{fig:ExclusionSensitivity}.
For all \bi\ values and all live times, the FWHM of $5$~keV yields a $\sim45\%$ higher sensitivity
with respect to a $10$~keV resolution.
The median sensitivity after $3$~months and $5$~years of data collection
in the two considered cases are reported in Tab.~\ref{tab:Sensitivity}.
The dependence of the median sensitivity on live time is typical of a background-dominated experiments:
namely, \cuore\ expects about one event every four days in a $\pm3\sigma$ region around \Qbb.
The results in Tab.~\ref{tab:Sensitivity} show also the importance of energy resolution and suggest to put
a strong effort in its optimization.
As a cross check, we compare the sensitivity just obtained with that provided by the analytical method
presented in~\cite{Alessandria:2011rc} and shown in dark green in Fig.~\ref{fig:ExclusionSensitivity}.
The analytical method yields a slightly higher sentitivity for short live times,
while the two techniques agree when the data sample is bigger.
We justify this with the fact that the uncertainty on the number of background counts
obtained with the Bayesian fit is slightly larger than the corresponding Poisson uncertainty
assumed in the analytical approach~\cite{Cowan:2011an}, hence the limit on \Tonbb\
is systematically weaker\footnote{See the discussion of the pulls for a more detailed explanation.}.
The effect becomes less and less strong with increasing
data samples, i.e. with growing live time. With a resolution of $5$~keV, the difference goes
from $8\%$ after $3$~months to $<0.1\%$ after $5$~years, while for a $10$~keV FWHM
the difference is $\sim6\%$ after $3$~months and $4\%$ after $5$~years.
One remark has to be done concerning the values reported in~\cite{Alessandria:2011rc}:
there we gave a $90\%$~C.I. exclusion sensitivity of $9.3\cdot10^{25}$~yr with $5$~yr of live time.
This is $\sim5\%$ higher than the result presented here and is explained
by the use of a different total efficiency: $87.4\%$ in~\cite{Alessandria:2011rc} and $81.3\%$
in this work.

We then extract the exclusion sensitivity after dividing the crystals
in $4$ subsets, as described in Sec.~\ref{sec:ExperimentalParameters}.
The median exclusion sensitivity values after $3$~months and $5$~years of data collection
with one and $4$ subsets are reported in Tab.~\ref{tab:Sensitivity}.
The division in subsets yields only a small improvement (at the percent level) in median sensitivity.
Based on this results only, one would conclude that dividing the data into subsets with different \bi\
is not worth the effort. This conclusion is not always true, and strongly relies
on the exposure and \bi\ of the considered subsets. As an example, we repeated a toy analysis
assuming a \bi\ of $10^{\mbox{-}2}$~\ctsper, and with two subsets of equal exposure
and \bi\ $0.5\cdot10^{\mbox{-}2}$~\ctsper\ and $1.5\cdot10^{\mbox{-}2}$~\ctsper, respectively.
In this case, the division of the data in to two subsets yields a $\sim10\%$ improvement
after $5$~yr of data taking. Hence, the data subdivision is a viable option for the final analysis,
whose gain strongly depends on the experimental BI of each channel.
Similarly, we expect the \cuore\ bolometers to have different energy resolutions;
in \cuoreo, these ranged from $\sim3$~keV to $\sim20$~keV FWHM~\cite{Alduino:2016zrl}.
In the real \cuore\ analysis a further splitting of the data can be done
by grouping the channels with similar FWHM, or by keeping every channels separate.
At the present stage it is not possible to make reliable predictions for the FWHM distribution
among the crystals, so we assumed an average value (of $5$ or $10$~keV) throughout the whole work.

Ideally, the final \cuore\ \onbb\ decay analysis should be performed
keeping the spectra collected by each crystal separate,
additionally to the usual division of the data into data sets comprised by two calibration runs~\cite{Alfonso:2015wkao}.
Assuming an average frequency of one calibration per month,
the total number of energy spectra would be $\sim 6\cdot10^4$.
Assuming a different but stationary \bi\ for each crystal,
and using the same \Co\ rate for all crystals,
the fit model would have $\sim10^3$ parameters.
This represents a major obstacle for any existing implementation of the Metropolis-Hastings
or Gibbs sampling algorithm.
A possible way to address the problem might be the use of different algorithms,
e.g. nested sampling~\cite{Feroz:2008xx,Handley:2015fda},
or a partial analytical solution of the likelihood maximization.

\begin{table}[h]
  \caption{\small{Median exclusion sensitivity for different energy resolutions and different subset numbers.}}
  \label{tab:Sensitivity}
  \begin{tabular}{cccc}
    \toprule
    FWHM      & $N_d$ & $\hat{T}_{1/2}^{\,0\nu}$ at $0.25$~yr & $\hat{T}_{1/2}^{\,0\nu}$ at $5$~yr\\
    $[$keV$]$ &       &  [yr]                              & [yr] \\
        \midrule
    $5$  & $1$ & $1.7\cdot10^{25}$ & $8.9\cdot10^{25}$ \\
    $10$ & $1$ & $1.2\cdot10^{25}$ & $6.1\cdot10^{25}$ \\
    $5$  & $4$ & $1.8\cdot10^{25}$ & $9.1\cdot10^{25}$ \\
    $10$ & $4$ & $1.3\cdot10^{25}$ & $6.2\cdot10^{25}$ \\
    \bottomrule
  \end{tabular}
\end{table}

We perform two further cross-checks in order to investigate the relative importance
of the flat background and the \Co\ peak.
In the first scenario we set the \bi\ to zero, and do the same for the \Co\ rate in the second one.
In both cases, the data are not divided into subsets, and resolutions of $5$ and $10$~keV are considered.
With no flat background and a $5$~keV resolution, no \Co\ event leaks in the $\pm3\sigma$ region around \Qbb\
even after $5$~yr of measurement.
As a consequence, the $90\%$~CI limits are distributed on a very narrow band,
and the median sensitivity reaches $1.2\cdot10^{27}$~yr after $5$~yr of data collection.
On the contrary, if we assume a $10$~keV FWHM, some \Co\ events fall in the \onbb~decay
ROI from the very beginning of the data taking.
This results in a strong asymmetry of the sensitivity band.
In the second cross-check, we keep the \bi\ at $1.02\cdot10^{\mbox{-}2}$~\ctsper,
but set the \Co\ rate to zero. 
In both cases, the difference with respect to the standard scenario is below $1\%$.
We can conclude that the \Co\ peak with an initial rate of $0.428$~cts/(kg$\cdot$yr)
is not worrisome for a resolution of up to $10$~keV,
and that the lower sensitivity obtained with $10$~keV FWHM with respect to the $5$~keV case
is ascribable to the relative amplitude of $\lambda^{bkg}_{di}$ and $\lambda^{0\nu}_{di}$ only
(Eqs.~\ref{eq:bkgLambda2} and~\ref{eq:onbbLambda}).
This is also confirmed by the computation of the sensitivity for the optimistic scenario
without the $1.9$~keV shift of the \Co\ peak used in the standard case.

We test the fit correctness and bias computing the pulls, i.e. the normalized residuals,
of the number of counts assigned to each of the fit components.
Denoting with $N^{bkg}$ and $N^{Co}$ the number of generated background and \Co\ events, respectively,
and with $M^{bkg}$ and $M^{Co}$ the corresponding number of reconstructed events,
the pulls are defined as:
\begin{equation}
  r_{bkg(Co)} = \frac{ M^{bkg(Co)}-N^{bkg(Co)} }{ \sigma_{M^{bkg(Co)}} }\ ,
\end{equation}
where $\sigma_{M^{bkg(Co)}}$ is the statistical uncertainty on $M^{bkg(Co)}$ given by the fit.

For an unbiased fit, the distribution of the pulls is expected to be Gaussian
with a unitary root mean square (RMS).
In the case of exclusion sensitivity, we obtain $r_{bkg}=-0.2\pm0.4$ and $r_{Co}=0.1\pm0.5$ for all live times.
The fact that the pull distributions are slightly shifted indicates the presence of a bias.
Its origin lies in the Bayesian nature of the fit and is that all fit contributions are constrained to be greater than zero.
We perform a cross-check, by extending the range of all parameters ($BI$, $R^{Co}$ and $\Gamma^{0\nu}$) to negative values.
Under this condition, the bias disappears.
In addition to this, an explanation is needed for the small RMS of the pull distributions.
This is mainly due to two effects: first, the toy-MC spectra are generated using $H_0$,
while the fit is performed using $H_1$; second, the statistical uncertainties on all parameters
are larger than the Poisson uncertainty on the number of generated events.
To confirm the first statement, we repeat the fit using $H_0$ instead of $H_1$
and we obtain pulls with zero mean and an RMS $\sim0.8$, which is closer to the expected value.
Finally, we compare the parameter uncertainty obtained from the fit
with the Poisson uncertainty for the equivalent number of counts,
and we find that the difference is of $O(20\%)$.

\begin{figure}[t]
  \def\svgwidth{\columnwidth}
  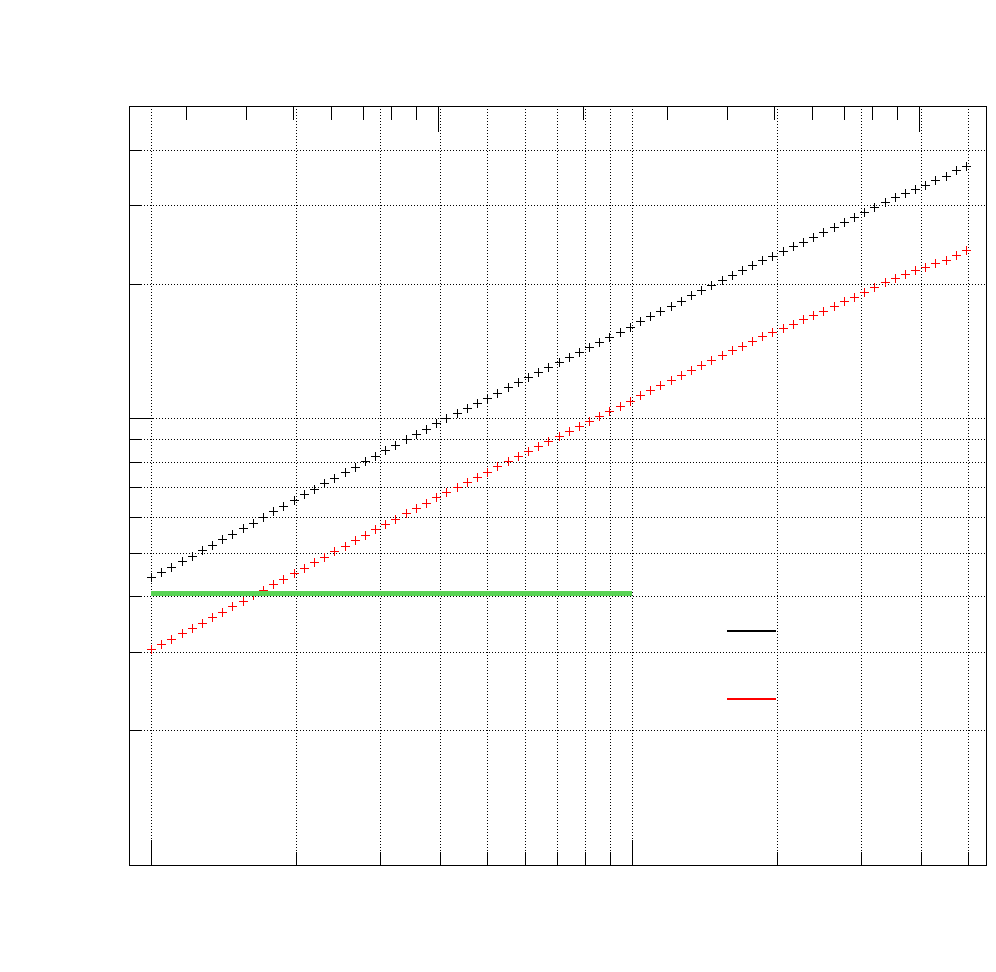
  \caption{\small{$3\sigma$ discovery sensitivity with a \bi\ of $1.02\cdot10^{\mbox{-}2}$ \ctsper\
      and an FWHM of $5$~keV.}}
  \label{fig:discoverysensitivity}
\end{figure}

\subsection{Discovery Sensitivity}\label{sec:DiscoverySensitivityResults}

The extraction of the discovery sensitivity involves
fits with the background-only and the background-plus-signal models.
Moreover, two multi-dimensional integrations have to be performed for each toy-MC spectrum,
and a loop over the \onbb\ decay half-life has to be done
until the condition of Eq.~\ref{eq:discoverySensitivity} is met.
Due to the high computation cost, we compute the $3~\sigma$ discovery sensitivity
for a FWHM of $5$ and $10$~keV with no crystal subdivision.
As shown in Fig.~\ref{fig:discoverysensitivity},
with a $5$~keV energy resolution \cuore\ has a $3~\sigma$ discovery sensitivity
superior to the limit obtained from the combined analysis of \textsc{Cuore}-0
and \textsc{Cuoricino} data~\cite{Alfonso:2015wkao} after less than one month of operation,
and reaches $3.7\cdot10^{25}$~yr with $5$~yr of live time.

Also in this case, the pulls are characterized by an RMS smaller than expected,
but no bias is present due to the use of $H_1$ for both the generation
and the fit of the toy-MC spectra.

\section{Conclusion and Outlook}\label{sec:Conclusion}

We implemented a toy-MC method for the computation of the exclusion
and discovery sensitivity of \cuore\ using a Bayesian analysis.
We have highlighted the influence of the \bi\ and energy resolution on the exclusion sensitivity,
showing how the achievement of the expected $5$~keV FWHM is desirable.
Additionally, we have shown how the division of the data into subsets
with different \bi\ could yield an improvement in exclusion sensitivity.

Once the \cuore\ data collection starts
and the experimental parameters are available,
the sensitivity study can be repeated in a more detailed way.
As an example, non-Gaussian spectral shapes
for the \onbb\ decay and \Co\ peaks can be used,
and the systematics of the energy reconstruction can be included.

\section*{Acknowledgments}

The \cuore\ Collaboration thanks the directors and staff of the Laboratori Nazionali del Gran Sasso
and the technical staff of our laboratories. CUORE is supported by
The Istituto Nazionale di Fisica Nucleare (INFN);
The National Science Foundation under Grant Nos. NSF-PHY-0605119, NSF-PHY-0500337, NSF-\\
PHY-0855314, NSF-PHY-0902171,
NSF-PHY-0969852, NSF-PHY-1307204, NSF-PHY-1314881, NSF-PHY-\\
1401832, and NSF-PHY-1404205; The Alfred P. Sloan Foundation;
The University of Wisconsin Foundation;
Yale University;
The US Department of Energy (DOE) Office of Science under Contract Nos. DE-AC02-05CH1-1231,
DE-AC52-07NA27344, and DE-SC0012654;
The DOE Office of Science, Office of Nuclear Physics under Contract Nos. DE-FG02-08ER41551 and DE-FG03-00ER41138;
The National Energy Research Scientific Computing Center (NERSC).

\bibliographystyle{hphysrev-5}
\bibliography{CuoreSensitivity}

\end{document}